\documentclass[manuscript]{aastex}

\begin{document}


\title{A Population of Intergalactic Supernovae in Galaxy Clusters}


\author{Avishay Gal-Yam\altaffilmark{1}, Dan Maoz}
\affil{School of Physics \& Astronomy, Tel-Aviv University, Tel-Aviv
69978, Israel}
\email{avishay@wise.tau.ac.il; dani@wise.tau.ac.il}

\author{Puragra Guhathakurta\altaffilmark{2}}
\affil{Herzberg Fellow, Herzberg Institute of Astrophysics, National 
Research Council of Canada, 5071 West Saanich Road, Victoria, B.C. V9E 
2E7, Canada}
\email{raja@ucolick.org}

\and

\author{Alexei V. Filippenko}
\affil{Department of Astronomy, 601 Campbell Hall, University of
California, Berkeley, CA 94720-3411, USA}
\email{alex@astro.berkeley.edu}


\altaffiltext{1}{Colton Fellow.}
\altaffiltext{2}{Permanent address: UCO/Lick Observatory, Department 
of Astronomy and Astrophysics, University of California, Santa Cruz, 1156 
High Street, Santa Cruz, CA 95064, USA}


\begin{abstract}

We have discovered seven type Ia cluster supernovae (SNe) in the course of the
Wise Observatory Optical Transients Search in the fields of $0.06<z<0.2$ galaxy
clusters. Two of these events, SN 1998fc in Abell 403 $(z=0.10)$ and SN 2001al
in Abell 2122/4 $(z = 0.066)$, have no obvious hosts. Both events appear 
projected on the halos of the central cD galaxies, but have velocity
offsets of 750-2000 km s$^{-1}$ relative to those galaxies, suggesting they
are not bound to them. We use deep Keck imaging
of the locations of the two SNe to put upper limits on the luminosities of
possible dwarf hosts, $M_R > -14$ mag for SN 1998fc and $M_R > -11.8$ mag for
SN 2001al. The fractions of the cluster luminosities in dwarf galaxies fainter
than our limits are $<3\times10^{-3}$ and $<3 \times 10^{-4}$,
respectively. Thus, $2/7$ of the SNe would be associated with
$\le3\times10^{-3}$ of the luminosity attributed to galaxies. 
We argue, instead, that the progenitors of both
events were probably members of a diffuse population of intergalactic stars,
recently detected in local clusters via planetary nebulae and red giants.
Considering the higher detectability of hostless SNe compared to normal SNe, we
estimate that $20^{+12}_{-15}$ percent of the SN~Ia parent stellar population
in clusters is intergalactic. This fraction is consistent with other
measurements of the intergalactic stellar population, and implies that the
process that produces intergalactic stars (e.g., tidal disruption of cluster
dwarfs) does not disrupt or enhance significantly the SN~Ia formation
mechanism. Hostless SNe are potentially powerful tracers of the formation of
the intergalactic stellar population out to high redshift.\\

\end{abstract}




\section{Introduction}

The existence of a diffuse population of intergalactic stars in galaxy clusters
was first discussed by Zwicky (1951), who claimed to measure excess starlight
between the galaxies in the core of the Coma cluster. This early result was
later debated (e.g., Welch \& Sastry 1971; Melnick, Hoessel, \& White 1977) and
was finally confirmed by recent deep CCD-based surveys (e.g., Feldmeier et al.
2002). The diffuse intergalactic population in the Coma cluster is also
manifest in large, low surface brightness arc-like features (Gregg \& West
1998; Trentham \& Mobasher 1998) that are also seen in other nearby clusters
(Calc{\' a}neo-Rold{\' a}n et al. 2000).
 
Individual intergalactic red giant (RG) stars have been detected with the
{\it Hubble Space Telescope (HST)} in the Virgo cluster (Ferguson, Tanvir, \&
von Hippel 1998; Durrell et al. 2002). Ground-based observations have revealed
intergalactic planetary nebulae (PNe) in the Virgo and Fornax clusters
(Arnaboldi et al. 1996; Theuns \& Warren 1997; Mendez et al. 1997; Ciardullo et
al. 1998; Feldmeier, Ciardullo, \& Jacoby 1998). Although some of the initial
candidate PNe turned out to be distant background objects (Kudritzki et
al. 2000), the majority of the discovered objects are indeed intergalactic PNe
(Ciardullo et al. 2002). Typically, these studies find that 10--20\% of the
stars in galaxy clusters are in the diffuse component. Preliminary results
suggest that intergalactic globular clusters may also be used as tracers of the 
intergalactic stellar population (Hilker 2002).  

The intergalactic population of stars is thought to be the result of the tidal
disruption of cluster galaxies, interacting with other cluster members or with
the cluster core potential (Richstone \& Malumuth 1983; Merritt 1984,
1985). Tidal disruption of cluster galaxies is reproduced in detailed
simulations of galaxy clusters (Dubinski, Mihos, \& Hernquist 1996; Moore et
al. 1996; Korchagin, Tsuchiya, \& Miyama 2001).  Recent simulations of both
rich and poor clusters produce a diffuse population containing about 10$\%$ of
the stars in the cluster. The spatial distribution of these stars can be
approximated by a continuation of the de Vaucouleurs profile of the central cD
galaxy out to a distance of $\sim1$ Mpc from the cluster center (Dubinski
1998). Tidal debris arcs, similar to those observed, are also reproduced (Moore
et al. 1996). We collectively refer to all the observed stellar components that
are not associated with a particular galaxy, 
including diffuse intracluster light, tidal
debris, and low surface brightness features, as the intergalactic stellar
population. We follow Feldmeier et al. (2002) and include the extended
envelopes of the central cD galaxies in this category, as it is observationally
unclear whether such envelopes are a feature of cD galaxies or a concentration
of intergalactic stars that settle at the bottom of the cluster potential
wells. Numerical simulations of rich clusters (Dubinski 1999) seem to confirm
the notion that extended cD envelopes and intergalactic tidally stripped stars
are one and the same.
        
Type Ia supernovae (SNe~Ia) are known to occur in galaxies of all classes and
morphologies. It is therefore plausible that such events also occur among the
intergalactic stellar population. Indeed, it is widely accepted that SN~Ia
progenitors are accreting or merging white dwarfs (WDs; see, e.g., Livio 2001,
and references therein). The observation of intergalactic RGs and PNe makes it
unavoidable that intergalactic WDs exist also, since they are the end products
of the RG-PN-WD evolutionary track.  The tidal forces presumably responsible
for removing these stars from galaxies are orders of magnitude too weak to
separate close binaries, so intergalactic WDs should also be able to accrete
from, or merge with, close companions, and then explode as SNe~Ia.

{\it Bona fide} intergalactic SNe have not been confirmed to date. An
interesting candidate, SN 1980I in the Virgo cluster (Smith 1981), is projected
on a stellar ``bridge'' between two nearby galaxies, and may be associated
with either of these hosts. Several other candidates have been proposed over
the years. In a search for SNe in nearby Abell clusters, Reiss et al.  (1998a)
detected two events with no visible host galaxy, SN 1998bt in Abell 1736
(Germany 1998) and SN 1998dy in Abell 3266 (Reiss 1998b).  A much deeper search
in duplicate {\it HST} images of distant galaxy clusters (Gal-Yam, Maoz, \&
Sharon 2002) revealed a possibly hostless SN in the $z=0.9$ cluster
MS1054.4-0321 (Gal-Yam, Sharon, \& Maoz 2001).  However, the lack of a
spectroscopic redshift for any of these candidates precludes confirmation of
their cluster membership.

In this paper, we report the discovery of two likely 
intergalactic SNe among a sample
of seven confirmed cluster SNe. Where applicable, we assume a flat universe
with a cosmological constant $\Omega_{\Lambda}=0.7$ and $h=0.7$, where $h$ is
the Hubble constant in units of 100 km s$^{-1}$ Mpc$^{-1}$.




\section{Observations}

The SNe discussed in this work were discovered by the Wise Observatory Optical
Transient Search (WOOTS). Full details about the design and results of this
program are given elsewhere (Gal-Yam \& Maoz 1999a; A. Gal-Yam \& D. Maoz, 
in preparation). Briefly, WOOTS is an imaging survey of a sample of 163 rich
galaxy clusters at redshifts $0.06\le z \le 0.2$.  The clusters were observed
in the years 1998--2001 with the Wise Observatory 1-m telescope. Using
unfiltered (``clear'') 600-s CCD images, WOOTS was sensitive to SNe with
magnitudes $R \lesssim 21.5$ mag. Transient or variable objects were detected
by comparing each new WOOTS frame to older templates, using
point-spread-function matching and image subtraction. An effort was made to
follow-up all transient events photometrically (with the Wise 1~m) and
spectroscopically (at various larger telescopes).  A total of 14 candidate SNe
were discovered in cluster fields, 11 of which were spectroscopically
confirmed. Seven of the confirmed events were SNe~Ia at the redshifts of their
respective galaxy clusters. These cluster SNe are listed in Table 1.  No
obvious hosts are detected in the Wise images of two of the cluster SNe,
1998fc and 2001al.

\subsection{SN 1998fc in Abell 403}

SN 1998fc was discovered in unfiltered Wise 1~m images of Abell 403, obtained
on December 20, 1998 (UT dates are used throughout this paper), and confirmed
on subsequent frames obtained on January 7 and 9, 1999 (Figure 1, upper left;
typical seeing $2.5''$).
No trace of this SN is evident in prior WOOTS images up to November 14, 1998.
Assuming the SN was discovered close to maximum light (see also below), we
have used the methods of Poznanski et al. (2002) to calculate the colors of
this SN at the time of our observations. Using calibrated CCD observations
of this field, kindly supplied by R. Gal and the DPOSS team, we selected
stars with similar colors, and used them to calculate the transformation
between unfiltered flux and Cousins $R$. The measured unfiltered flux translates 
to $R=20.5$ mag (see Gal-Yam \& Maoz 1999b, and Table 1).
No hint for an underlying host is seen in the deepest WOOTS images of
this field, obtained prior to the explosion of SN 1998fc.
Figure 1 shows the neighborhood of SN 1998fc. 

We determined the effective radii of galaxies by fitting both the Wise images
and additional Keck imaging (see \S~2.3 below), having $\sim2''$ seeing
for Abell 403 and $\sim0.5''$ seeing for Abell 2124/2,
with a Sersic (1968) profile, using the galaxy-fitting
package GALFIT (Peng et al. 2002). We allowed the Sersic index $1/n$ to 
vary, including the cases of a de Vaucouleurs (1948) profile $(n=4)$ and
an exponential disk $(n=1)$. The effective radius values quoted hereafter
are for the best-fitting Sersic indices.
The galaxy projected closest to the SN is labeled No. 1. 
However, since the SN is located $14.8''$ away from
the nucleus of this galaxy, more than 6 effective radii, 
it is unlikely to be related to it.
In fact, the only viable host galaxy of SN 1998fc,
considering the effective radii of all nearby objects detected in WOOTS images,
is the cD galaxy of Abell~403 (PGC 011298, Paturel et al. 1989), projected
$26.8''$ from the SN ($40h^{-1}$ kpc). We will argue below, however, that the 
SN and the cD have quite different radial velocities. 

A spectrum of SN 1998fc was obtained on Jan. 14, 1999, with the EFOSC
spectrograph mounted on the ESO 3.6m telescope (Gal-Yam \& Maoz 1999c).
Another spectrum was obtained on Jan. 20, 1999, with the Low Resolution Imaging
Spectrometer (LRIS; Oke et al. 1995) mounted on the Keck-II 10-m telescope
(Filippenko, Leonard, \& Riess 1999). The LRIS spectrum, which is of superior 
quality, was not obtained near the parallactic angle, but in view of the low
airmass (1.10) during the integration we expect very little atmospheric 
dispersion (Filippenko 1982). Initial inspection of both spectra showed that 
the SN is of type
Ia, about one month after maximum, and at approximately the same redshift 
as the cluster (Figure 1, bottom). Again, there are
no signs of a host galaxy in the SN spectra.

In order to investigate more carefully the association of the SN with the
cluster and with particular galaxies in the cluster, we have attempted to
measure accurately the redshift of the SN. However, there are several
complicating factors. First,
spectra of SNe~Ia at this age consist of broad blended features from 
which it is difficult to determine an accurate redshift. Second, the ejecta
of SNe Ia of a given age have a range of velocities, as estimated from
the blueshifted centroids of various absorption lines. In general, the lines
become less blueshifted with time, as more slowly expanding gas at inner radii
becomes optically thin. Thus, reliably estimating a SN velocity relative to a
galaxy requires a good knowledge of the age of the SN and of the scatter in
measured SN ejecta velocities for that age. Part of that scatter will also be
due to the orbit of the SN in the potential of the host galaxy.

To establish the plausible range of ages for SN 1998fc at the time the LRIS
spectrum was obtained, we have compared its spectral features to those of high
signal-to-noise ratio (S/N) spectra of nearby SNe~Ia at similar ages, drawn 
from the spectral archive presented by Poznanski et al. (2002). This comparison,
and the rest of the analysis described below, was done after we divided both
new and archival spectra by a second-order polynomial fit to the continuum.
This procedure assures that our analysis is not affected by low-frequency
distortions in the continuum shape, possibly due to atmospheric dispersion
or imperfect spectral response calibration. From this
comparison we can establish an age of 21 to 42 days past $B$-band maximum.
Younger or older ages are ruled out by the spectra, as visual inspection 
reveals strong differences in the relative strengths of the major spectral
features. Comparison of the Wise
photometry of this event to the light curves of SNe Ia (Riess et al. 1999)
raises the lower limit on the age to $>25$ days. 

Next, we created a composite
spectrum of seven spectra of SNe~Ia at $28-42$ days past maximum light,
drawn from the archive of Poznanski et al. (2002).
The individual spectra were first deredshifted according to their host-galaxy
redshifts. The composite spectrum
was created by forming the mean of all spectra, cross-correlating each
individual spectrum against the mean, and shifting the spectrum to zero
velocity relative to the mean. A new mean spectrum was then calculated and the
process repeated iteratively several times. 
We cross-correlated each of the original seven
spectra of SNe~Ia mentioned above with the final template. Among this
group of spectra, one can see both the trend of velocity rising with time
(i.e.,  decreasing blueshift), and the 
scatter in velocity among several SNe at a  given age.
The highest velocity measured relative to the template 
is $\sim360$ km s$^{-1}$, and the standard 
deviation is $\sim190$ km s$^{-1}$. The template spectrum
of a SN Ia in this age range therefore has a systematic 
zero-point velocity uncertainty
of $\sim190/{\sqrt{7}} \approx 72$ km s$^{-1}$. The uncertainty in the velocity 
derived for an observed SN of this age will then be 
$\pm 190$ km s$^{-1}$ (statistical) $\pm 72$ km s$^{-1}$ (systematic).
Additional random error will be caused by noise in the observed spectrum.   

We have cross correlated the LRIS spectrum of SN 1998fc with the 
template described above and
measure a redshift of $z = 0.1023 \pm 0.0010$ for SN 1998fc.  The uncertainty
was determined from Monte-Carlo simulations in which the composite spectrum was
repeatedly degraded in S/N to a degree similar to our LRIS spectrum, and
cross-correlated with the composite. To this random error we added the
systematic and statistical errors discussed above.
Struble \& Rood (1999) tabulate a redshift of $z = 0.1033$ for this cluster,
based on the average of two measurements of the cD galaxy's redshift, 
one obtained by Shectman (1985)
with $z=0.0996\pm0.00017$, and the second by Kristian, Sandage, 
\& Westphal (1978) with $z=0.107$. We have ascertained (S. Shectman, 2002, private
communication) that both measurements are 
of the same galaxy, so a discrepancy of 2000 km s$^{-1}$ exists between these 
two redshift determinations. In either case, our measured redshift confirms 
the cluster membership of the SN.

Let us consider the measurement by Shectman (1985) of $z=0.0996$ for 
PGC 011298, the cD galaxy which is the only plausible bright galaxy that 
could host SN 1998fc. The velocity difference between PGC 011292 and 
SN 1998fc is $\sim750$ km s$^{-1}$. Given the mean scatter we have
measured above and Shectman's measurement accuracy ($50$ km s$^{-1}$), 
this velocity offset is significant. It suggests that SN 1998fc may not be
bound to this galaxy. Note that if the measurement by Shectman (1985) 
is erroneous, and that of Kristian et al. (1978) is correct, the velocity
difference is even larger ($\sim1300$ km s$^{-1}$). Assuming 
the redshift accuracy given by Sandage, Kristian, \& Westphal (1976), 
$\Delta z=0.001$, a significant velocity offset between that galaxy and 
SN 1998fc will still exist, and the same conclusion will hold. A confirmation 
of the redshift of PGC~011298 could further illuminate this issue.

\subsection{SN 2001al in Abell 2122/4}

SN 2001al was discovered in unfiltered Wise 1-m images of Abell 2122/4 obtained
on March 26, 2001, and confirmed on unfiltered and $R$-band frames obtained on March
28 (Figure 2, upper left; typical seeing $\sim2''$). 
>From these, we measure $R=21.4$ mag. No trace of
this SN is evident in prior WOOTS images, obtained on August 20, 2000 (see
Gal-Yam \& Maoz 2001, and Table 1). 
A search for possible underlying host galaxies in prior WOOTS images of this
field reveals no likely candidates. Figure 2 shows the location of SN 2001al
and nearby galaxies. The closest galaxies detected near SN 2001al are marked as
galaxies No. 1 and 2. The SN is located $8.4''$ and $8.5''$ away from the
nuclei of these galaxies, which is 40 and 6 times their effective radii, 
respectively. The cD
galaxy of A2122/4, UGC 10012, is projected $112h^{-1}$ kpc away from SN 2001al.

Prompt follow-up spectroscopy obtained on
March 29, 2001, with LRIS mounted on the Keck-I 10-m telescope (Filippenko,
Barth, \& Leonard 2001), shows that the SN is of type Ia, several weeks after
maximum (Figure 2, bottom). The spectrum was obtained at low airmass
(1.1), near the parallactic angle. 
Using the same methods as for SN 1998fc, 
we estimate, based on comparison to our spectral database,
that the spectrum was obtained 21 to 33 days after maximum light. 
Cross-correlation against a template composed of nearby SN Ia 
spectra that are similar to SN 2001al and in this age bracket gives a   
redshift of $0.0723 \pm 0.0011$. We note, however, that the agreement
between the spectrum of SN 2001al and the SN~Ia template spectrum is not as
good as for SN 1998fc. This may be the result of some problem in the
SN 2001al data, or a real spectral peculiarity in this SN. 
This disagreement may affect the accuracy of the redshift
determined by cross-correlation. A visual ``chi-by-eye''
comparison of the LRIS spectrum with
redshifted versions of the template allows us to set limits on the allowed
range of redshifts for this event, $0.070\le z \le0.076$, which is within
$\sim 1100-2800$ km s$^{-1}$ of the redshift\footnotemark[1] of Abell 2122/4
($z=0.0661$; Struble \& Rood 1999). 
As in the case of SN 1998fc, there appears to be a large velocity difference
between SN 2001al and the cD galaxy UGC 10012 ($\sim1700$ km s$^{-1}$). 
This suggests the SN is not
bound to the cD galaxy, even if we consider the uncertainties in the 
SN redshift determination, as discussed above.

\footnotetext[1]{Abell, Corwin, \& Olowin (1989) list two clusters less than
$5'$ away from the location of SN 2001al, Abell 2122 and 2124.  Struble \&
Rood (1999) give the same redshift for both clusters ($z = 0.0661$, based on 7
and 63 measurements for A2122 and A2124, respectively), which is also the
redshift of the cD galaxy, UGC 10012. Thus it appears that all available data
are best explained by one cluster, at $z = 0.0661$, which we denote by A2122/4,
with the cD galaxy UGC 10012 at its center.}

\subsection{Deep Imaging of the SN Locations}

The lack of visible host galaxies in WOOTS imaging suggests that SNe 1998fc and
2001al may be the first examples of intergalactic SNe. However, one must
consider the possibility that these SNe reside in faint dwarf galaxies,
invisible in our WOOTS images. An illustrative example is the case of the
SN Ia 1999aw, whose faint host has absolute magnitudes $M_B=-12.2$,
$M_V=-12.5$, and $M_I=-12.2$ mag (Strolger et al. 2002).  To test for the possible
existence of such faint dwarf host galaxies, we undertook deep imaging of the
locations of SNe 1998fc and 2001al.

We observed the location of SN 1998fc on January 18, 2002, using LRIS mounted
on the Keck-I 10~m telescope. Sixteen 60-s frames were obtained in direct
imaging mode, with no filter. Observing conditions were poor, with $\sim2''$
seeing and imminent fog. Nevertheless, tying the combined 960-sec unfiltered
image with our calibrated WOOTS frames, we estimate that the limiting magnitude
for point sources of the unfiltered LRIS image is equivalent to 
$R=24.5$ mag (with $2''$ seeing, typical dwarf galaxies at the cluster redshift
would be practically unresolved). No object is seen at or near
the location of SN 1998fc down to the limiting magnitude of the LRIS image
(Figure 1, upper right). This translates to a limit of $M_R > -14$ mag for any
possible dwarf hosts.

On March 11, 2002, we observed the location of SN 2001al using ESI (Sheinis et
al. 2002) mounted on the Keck-II 10-m telescope. Exposures totaling 600 s in
$B$, 600 s in $I$, and 200 s in $R$ were obtained under good conditions, with
$0.7''$ seeing. A possible faint source was marginally detected in the $I$-band and
$B$-band images. To further check the reality of this source we inspected a
deep 8400 s $R$-band image of this field obtained with LRIS in 1997, kindly
made available by J. Blakeslee (see Blakeslee \& Metzger 1999, for
details). This image has limiting surface brightness $\sim26$ mag
arcsec$^{-2}$, and seeing $\sim0.5''$. The possible source detected in our 
ESI images is resolved in the deeper LRIS $R$-band image into three
faint sources, located near the position of SN 2001al (Figure 2, upper
right). Tying the LRIS $R$ band with our calibrated WOOTS photometry, we
measure $R = 26.0 \pm 0.2$ mag for the brightest of these sources (marked as
``Source 1''). At the redshift of SN 2001al ($z=0.0723$), 
we find an absolute magnitude of $M_R = -11.8$ for this
possible host galaxy. Source 1 is $\sim1''$ from the location of SN 2001al.

However, at such faint magnitudes, the probability of finding unrelated
background sources near any given location is significant. From the deep
$R$-band image, we measure a surface density of 0.02 sources per square arc
second, that are as bright as or brighter than Source 1, in this area.  The
probability of finding such a source within $1''$ of any location is
significant, $P = 0.06$. Nevertheless, in what follows we will explore the
consequences of assuming that one of the faint sources is the SN host.

\section{Discussion and Conclusions}

Using deep images of the locations of SNe 1998fc and 2001al we have set upper
limits on the luminosity of possible dwarf host galaxies. We now address the
question of whether or not these SNe are part of the intergalactic stellar
population, by considering the galaxy luminosity function in clusters. It is
well documented that SN rates are roughly proportional to host-galaxy
luminosity (e.g., Tammann 1970; van den Bergh \& Tammann 1991).
Given the galaxy luminosity
function in clusters, one can calculate the fraction of a cluster's luminosity
that is contained in dwarfs that are fainter than a certain limit, and thus
estimate the probability that a SN is associated with such faint hosts.

Recently, Trentham \& Tully (2002) measured the luminosity function in the
Virgo cluster down to very low absolute magnitudes ($M_R = -10$ mag). They find
that their measurements in the $R$-band are well fit by a modified Schechter
function,

\begin{equation}
N(M)dM = N_g \, {e}^{-{{(M - M_g)^2/(2\sigma_g^2)}}} + N_d \, 
(10^{[-0.4(M-M_d)]})^{\alpha_d+1} \, {e}^{-{10^{[-0.4(M-M_d)]}}} dM,  
\end{equation}

\noindent
where the contribution of giant galaxies is parameterized by a Gaussian with
normalization $N_g=17.6$, a characteristic peak magnitude $M_g=-19.5$ mag, and a
dispersion $\sigma_g=1.6$ mag. Dwarf galaxies are parameterized by a Schechter
function with an exponential cutoff magnitude $M_d=-18$, normalization
$N_d=66.88$, and a faint-end slope $\alpha_d=-1.03$. Integrating this function,
we derive the fraction $f$ of the luminosity contained within dwarf galaxies
that are fainter than some minimum luminosity $L_{min}$,

\begin{equation}
f(<L_{min}) = {\int^{L_{min}}_{0} N(L) L dL \over L_{total}} .
\end{equation}

\noindent
After converting the absolute magnitudes given by Trentham \& Tully from
$h=0.77$, implied by their use of the distances to Virgo galaxies taken from
Tonry et al. (2001), to $h=0.7$, we find $f(<L_{min})=0.0024$ and $0.0003$ 
for luminosity values corresponding to limiting absolute magnitudes 
$M_{R}=-14$ and $M_{R}=-11.8$ mag, respectively. 

Assuming that the Trentham \& Tully luminosity function is representative of
rich clusters in general (such as the clusters in the WOOTS sample), and that
the SN rate per galaxy is proportional to its luminosity, we have performed a
Monte-Carlo simulation to estimate the probability that two out of seven
cluster SNe would be associated with dwarf host galaxies fainter than the
limits obtained above. This was done by drawing $10^4$ ensembles of 7 host
galaxies with luminosities drawn from the luminosity-weighted distribution
function (the integrand in Eq. 2), and counting how many times two or more of
the host galaxies were as faint as, or fainter than, our limits.  We find this
probability is small, $\le6 \times 10^{-5}$, indicating it is unlikely that the
two SNe we have found belong to the stellar population bound to galaxies.

A possible caveat is that it is more difficult to discover SNe within bright
host galaxies than SNe in fainter, or undetected, hosts. The five WOOTS SNe
with detected hosts could constitute only a fraction of the true number of SNe
associated with bright hosts, while the majority of such events had remained
undetected.  The true fraction of SNe with very faint hosts would then be
smaller, and perhaps consistent with the luminosity function analysis above.
We have carried out detection efficiency experiments by adding artificial SNe
to real WOOTS images, with the SNe distributed in position like the starlight
(see Gal-Yam et al. 2002 for more details). We then processed the data like
real data, and noted the detection efficiency for such normal SNe, relative to
hostless SNe that are placed far from any galaxy.  Typical values are
$0.7~(0.4)$ for SNe with $R = 20.5~(21.5)$ mag. Taking these values into account,
our efficiency-corrected sample contains 10 events, 8 in bright hosts and 2
hostless. We find that the probability for 2/10 of the SNe to reside in
undetected dwarfs, calculated as above, is still small, $\le8 \times 10^{-5}$.

The most plausible conclusion is that SNe 1998fc and 2001al are indeed related
to the intergalactic stellar population. The two intergalactic SNe imply that
$20^{+12}_{-15}$ percent of the stellar population in clusters is
intergalactic, after accounting for the relative detection efficiencies between
normal and hostless SNe, and binomial statistics.

We note that there was no bias we were aware of that would favor successful
spectroscopic follow-up of either normal or hostless SNe among the 14 SN
candidates found by WOOTS in cluster fields. The 2/7 ratio should therefore
be representative of the true fraction we would have found with full
spectroscopic confirmation.  Among the three candidates lacking spectroscopy,
all appear to be associated with hosts. Even if there was a bias, and all three
are cluster SNe~Ia, the probability that SNe 1998fc and 2001al have dwarf hosts
is still $\le1 \times 10^{-4}$. The fraction of intergalactic 
SNe in our sample would then
be $20\%~(2/10)$, implying that $16^{+15}_{-10}$ percent of the stellar 
population in clusters is intergalactic, after accounting for our 
detection efficiency as above.

A final point we consider is the possible connection between SNe 1998fc and
2001al and the cD galaxies of their host clusters. The main argument suggesting
that the progenitors of these events were not associated with the cD galaxies 
is, in both cases, the large differences in velocity between the measured
velocities of the SNe and the values listed for the galaxies in the literature.
We note, however, that the discrepant redshifts of PGC
011298 (possibly hosting SN 1998fc) and the slight peculiarities in the 
spectrum of SN 2001al preclude a definite statement in both cases. 
It is conceivable that errors in the published redshifts of the cD galaxies 
have conspired with exceptionally
large deviations of the intrinsic velocities of these events (relative to the
mean for SNe~Ia of similar age) to produce large apparent differences in
redshift, when in fact these
SNe occurred in bound members of the galaxies. 

Even if this is the case, both events reside in the sparse outer halos of 
these giant galaxies, with SN 2001al projected 160 kpc (for $h=0.7$) from the
galaxy center. As discussed in $\S~1$, the extended envelopes of cD galaxies
are frequently considered a component of the intergalactic stellar population.
In other words, the distinction between the remote envelope of the central
galaxies in rich clusters and the intracluster stellar population may simply be
a matter of semantics. High-quality spectroscopy of a few more examples of
hostless SNe in galaxy clusters could resolve this 
question, by testing whether the redshift offsets between the SNe and the cD
galaxies in their host clusters are consistently larger than 
the expected intrinsic scatter in SN~Ia velocities.
If the model predictions of Dubinski (1998) are 
correct, and the intergalactic stellar population is distributed with
a de Vaucouleurs-like profile with an effective radius of order 1 Mpc, then
some intergalactic SNe will be discovered at very 
large projected distances from the central galaxies. 
Their intergalactic nature would then be unquestionable,
regardless of the subtleties of SN redshift determination.

Key questions in galaxy-cluster evolution include the evolution of the
early-to-late and blue-to-red galaxy fractions, cluster-induced onset and
turn-off of star formation in galaxies, and the possible transformation of
spirals into S0 galaxies (e.g., Butcher \& Oemler 1984; Dressler et al. 1997;
Couch et al. 1998; Poggianti et al. 1999).  These questions are potentially
connected with the evolution of the intergalactic stellar population, since the
same processes that affect cluster galaxies may also be responsible for the
creation of the intergalactic component. However, existing methods that probe
this intergalactic population are severely limited beyond the local
Universe. Measuring low surface brightness diffuse light is hindered by
cosmological $(1+z)^4$ surface-brightness dimming, and searches for individual
PNe or RGs are limited by the faintness of these objects. Intergalactic SNe~Ia,
being bright point sources, are visible to great distances and could be used as
tracers of the intergalactic stellar component out to high $z$. Admittedly,
limits on dwarf hosts would be increasingly difficult to obtain at higher
$z$. Still, possible hosts with absolute magnitudes as faint as $M_R \le -14$ mag
would have $R \le 28.3~(R \le 30.1)$ mag at $z = 0.5~(z = 1)$, so useful limits
could be set by observations with 8-10-m class ground-based telescopes, or with
{\it HST}. Numerical simulations (e.g., Dubinski 1998) show that the accessible
redshift range ($z \le 1$) covers the period of intense tidal stripping of
galaxies in rich clusters, suggesting the possibility of following the growth
of the intergalactic stellar population by measuring the intergalactic SN
fraction as a function of $z$.

To summarize, we have presented a sample of seven cluster SNe discovered by
WOOTS. Of these events, two have no obvious hosts other than the cluster 
cD galaxies in whose halos the SNe appear. We have estimated the intrinsic
redshifts of the hostless SNe, compared them to published redshifts
of the cDs, and argued that the SNe are unlikely to be bound to the cDs. 
Using deep Keck
images, we have set upper limits on the luminosity of possible dwarf host
galaxies. Applying the recently measured luminosity function of the Virgo
cluster to Abell 403 and Abell 2122/4, we have argued that the fraction of the
luminosity of these clusters contained in faint dwarfs is extremely small, so
that such galaxies are highly unlikely to host two out of the seven WOOTS
cluster SNe. This statement holds even when considering that we have missed
some of the SNe that occurred in bright hosts. We conclude that, most likely,
the progenitors of both hostless SNe were members of the intergalactic stellar
population, which we have detected by means of SNe for the first time. The
intergalactic stellar fraction that we find, $20^{+12}_{-15}$ percent, is
consistent with the fraction found by other means in nearby clusters. This
is in line with
the expectation that the process that produces intergalactic stars
(e.g., tidal disruption of cluster dwarfs) does not disrupt or enhance
significantly the SN~Ia formation mechanism. The existence of intergalactic
SNe needs to be confirmed by means of additional examples of the phenomenon.
Such SNe can serve as tracers of intergalactic stars in galaxy clusters out to
high $z$. The effects that a population of intergalactic SNe may have on the
properties of the intracluster medium (e.g. Sasaki 2001) 
remains a topic for further study.

\section*{Acknowledgments}

We are grateful to J. Blakeslee for kindly providing access to his deep LRIS
image of A2122/4. P. Leisy, O. Hainaut, and T. Sekiguchi are thanked for
obtaining a spectrum of SN 1998fc with the ESO 3.6m telescope, and the ESO and
La Silla directors for approving our ToO request. We acknowledge A. J. Barth,
D. C. Leonard, A. G. Riess, M. Geha, R. Chornock, C. Sorenson, and B. Schaefer
for their help with Keck spectroscopy and imaging, R. Stathakis for 
obtaining data at the AAT, R. Pogge for spectra obtained
with the KPNO 4-m Mayall telescope, S. Shectman for useful discussions,
R. Gal and the DPOSS project team for providing access to 
their CCD calibration library, and C. Peng for help with GALFIT. 
Special thanks go to E. O. Ofek, Y. Lipkin, D. Poznanski, and K. Sharon,
and to the Wise Observatory staff, especially J. Dann. AG and DM 
acknowledge support by the Israel Science Foundation --- the
Jack Adler Foundation for Space Research, Grant 63/01-1. AVF's work is supported
by NSF grant AST--9987438. This research has made use of the NASA/IPAC
Extragalactic Database (NED) which is operated by the Jet Propulsion Laboratory,
California Institute of Technology, under contract with the National Aeronautics
and Space Administration. We have also made use of the LEDA database
(http://leda.univ-lyon1.fr).

\clearpage


\begin{figure}
\centering
\caption{{\it Upper left:} A $90'' \times 90''$ section of the
WOOTS unfiltered discovery image of SN 1998fc in Abell 403.  The SN and nearby
galaxies are marked. North is up and east is to the left.  {\it
Bottom:} Observer-frame Keck/LRIS spectrum of SN 1998fc (light) compared 
with a composite
spectrum of local SNe~Ia redshifted to $z=0.1023$,
$\sim30$ days past maximum brightness (bold,
see text).
 {\it Upper right:} A $26''
\times 26''$ section of the Keck/LRIS image obtained in 2002 shows no object
within the SN error circle (smaller than the area at the center of the broken
``X'' mark) down to a limiting magnitude of $R \approx 24.5$ mag (see text).}
\end{figure} 

\begin{figure}
\centering
\caption{{\it Upper left:} A $3' \times 3'$ section of the 
WOOTS unfiltered discovery image of SN 2001al in Abell 2122/4. 
North is up and east is to the
left. {\it Bottom:} Observer-frame Keck/LRIS spectrum of SN 2001al (light) compared 
with a composite spectrum of local SNe~Ia
redshifted to $z = 0.0723$, $\sim30$ days past 
maximum (bold).
{\it Upper right:} A $25'' \times 25''$ section of the deep 
$R$-band Keck-LRIS image from Blakeslee \& Metzger (1999), centered on the
location of SN 2001al. Three faint sources are detected near the SN error 
circle (marked). For orientation purposes, 
Galaxies 1 and 2 from the left panel are also marked. 
}
\end{figure} 






\clearpage
\begin{deluxetable}{cccccc}
\tabletypesize{\scriptsize}
\tablecaption{Type Ia SNe discovered in galaxy clusters by WOOTS \label{SN table}}
\tablewidth{0pt}
\tablehead{
\colhead{SN} & \colhead{Cluster} & \colhead{Redshift} &
\multicolumn{2}{c}{Discovery} & \colhead{References\tablenotemark{a}}\\
\colhead{} & \colhead{} & \colhead{} & 
\colhead{Date} & \colhead{$R$ magnitude\tablenotemark{b}} & \colhead{}
}
\startdata
1998fc & Abell 403 & 0.10 & Dec. 20 1998 & 20.5 & $1,2,3$\\
2001al & Abell 2122/4 & 0.07 & Mar. 26 2001 & 21.4 & $4,5$\\
\tableline
1998eu & Abell 125 & 0.18 & Nov. 14 1998 & 20.7 & $1,6$\\
1999cg & Abell 1607 & 0.14 & Apr. 15 1999 & 20.0 & $7,8$\\
1999ch & Abell 2235 & 0.15 & May. 13 1999 & 19.8 & $7,9$\\
1999ci & Abell 1984 & 0.12 & May. 15 1999 & 20.4 & $7,9$\\
1999ct & Abell 1697 & 0.18 & Jun. 13 1999 & 21.2 & $10,11$\\
\enddata
\tablenotetext{a}{
References: (1) Gal-Yam \& Maoz 1999b, (2) Filippenko et al. 1999, (3) 
Gal-Yam \& Maoz 1999c, (4) Gal-Yam \& Maoz 2001, (5) Filippenko et al. 2001
(6) Gal-Yam \& Maoz 1998, (7) Gal-Yam \& Maoz 1999d, (8) Gal-Yam, Maoz, \&
Guhathakurta 1999, (9) Gal-Yam, Maoz, \& Pogge 1999, (10) Gal-Yam \& Maoz 
1999e, (11) Gal-Yam, Maoz, \& Guhathakurta 2000.
}
\tablenotetext{b}{~The magnitudes reported here involve improved 
subtraction of host galaxy light, and therefore supersede our previous
results, published in IAU Circulars.}
\end{deluxetable}

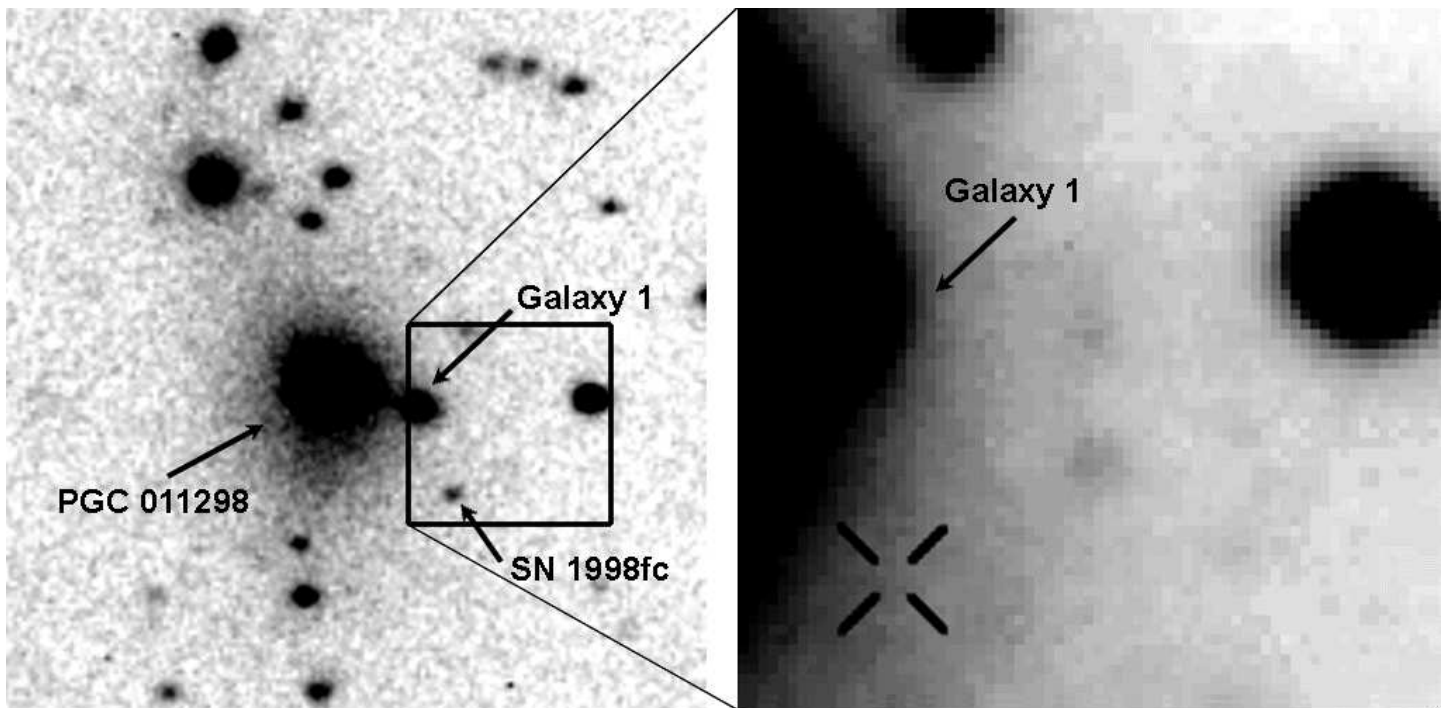
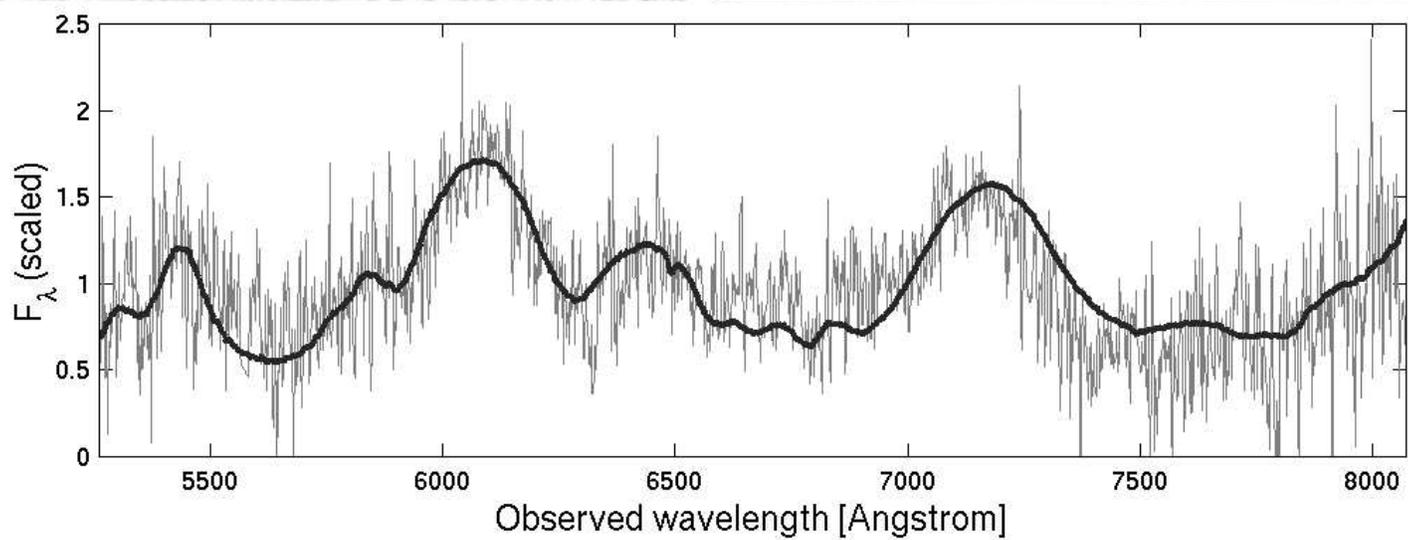
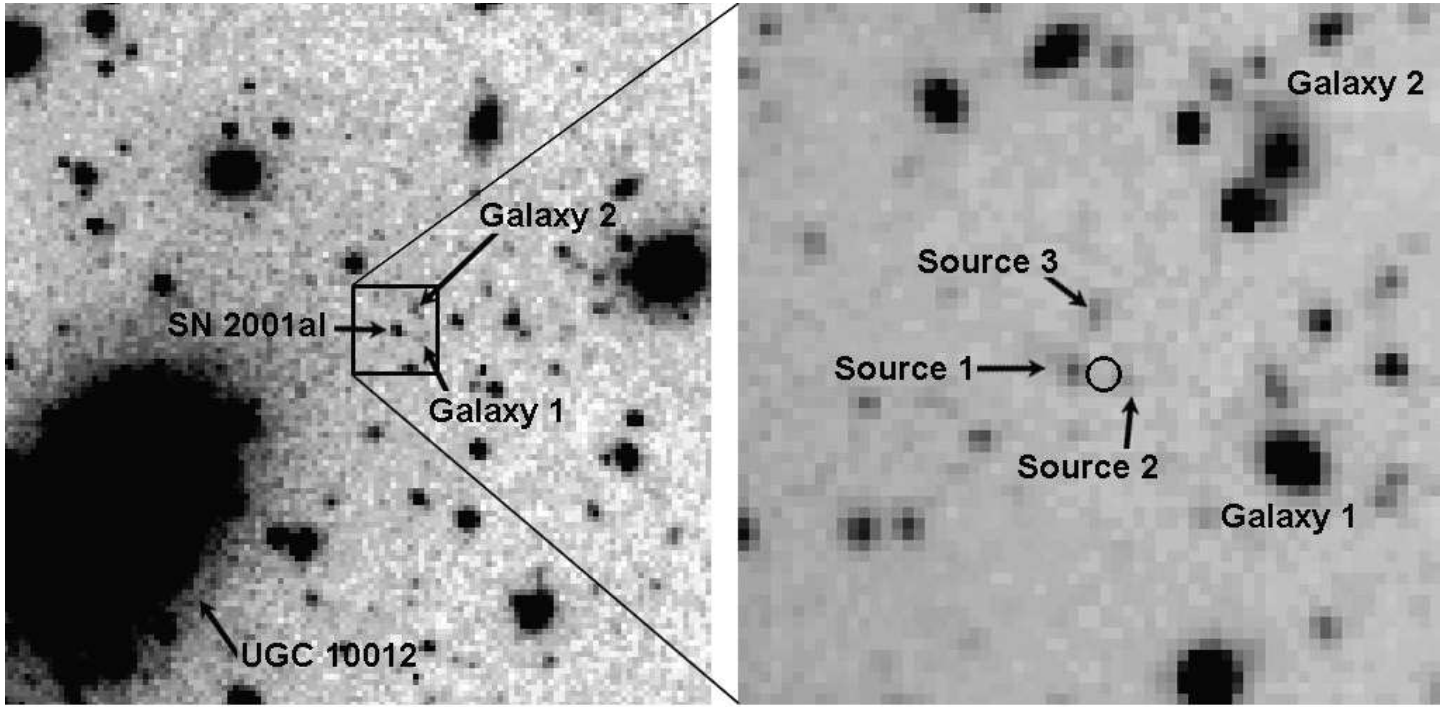
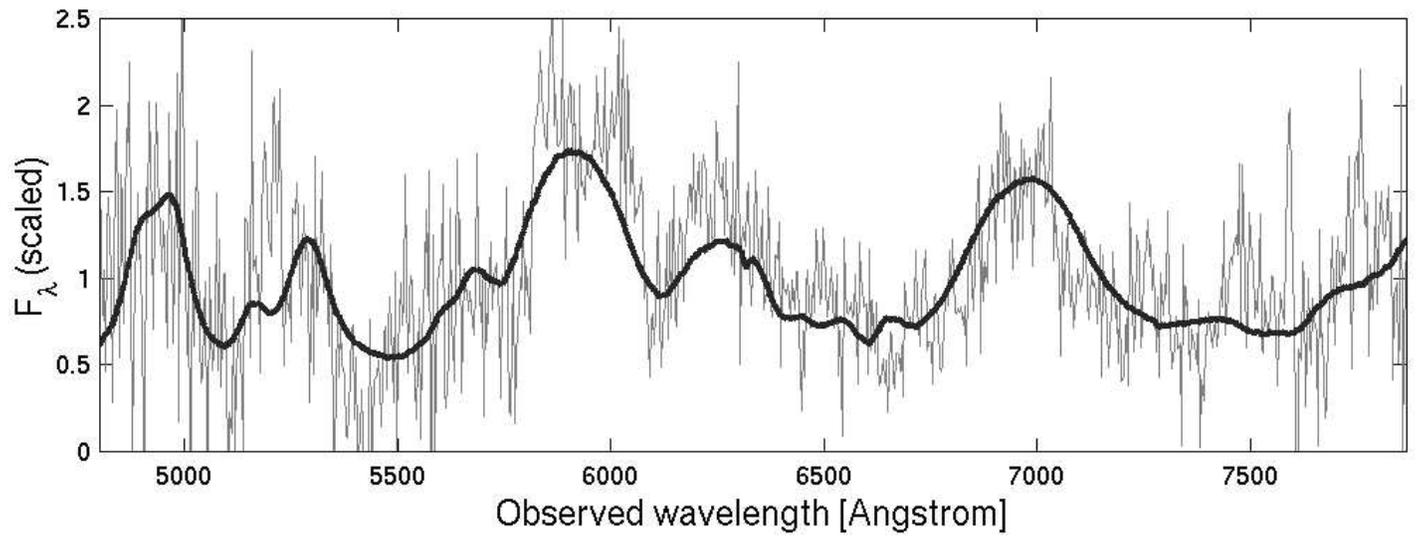
\end{document}